\documentclass[conference]{IEEEtran}
\IEEEoverridecommandlockouts
\usepackage{cite}
\usepackage{amsmath,amssymb,amsfonts}
\usepackage{algorithmic}
\usepackage{graphicx}
\usepackage{textcomp}
\usepackage{xcolor}
\usepackage[nolist]{acronym}
\usepackage{amsmath}
\usepackage{amssymb}
\usepackage{bbm}
\usepackage{booktabs}
\usepackage{relsize}
\usepackage{algorithm}
\usepackage{algorithmic}
\def\BibTeX{{\rm B\kern-.05em{\sc i\kern-.025em b}\kern-.08em
    T\kern-.1667em\lower.7ex\hbox{E}\kern-.125emX}}

\usepackage{pifont}
\newcommand{\cmark}{\ding{51}}
\newcommand{\xmark}{\ding{55}}

\usepackage{fancyhdr, lipsum}
\fancypagestyle{mahmood}{%
   \fancyhf{} 
   
   \fancyhead[C]{\copyright~2026 IEEE. Reprinting or republishing this material for the purpose of advertising or promotion, creating new collective works, reselling or redistributing to servers or lists, or using any copyrighted component in other works must adhere to IEEE policy. The paper has been accepted for publication in \textbf{IEEE SoftCOM 2026}.}
}%
\makeatletter
\let\ps@IEEEtitlepagestyle\ps@mahmood
\makeatother

\begin{document}


\begin{acronym}
    \acro{LLM}{Large Language Model}
    \acro{BS}{Base Station}
    \acro{ABS}{Aerial Base Station}
    \acro{UAV}{Unmanned Aerial Vehicle}
    \acro{AI}{Artificial Intelligence}
    \acro{QoS}{Quality of Service}
    \acro{QoE}{Quality of Experience}

    \acro{E2E}{End-to-End}
    \acro{UE}{User Equipment}
    \acro{LoS}{Line-of-Sight}
    \acro{MINLP}{Mixed Integer Non-linear Programming}
    \acro{DRL}{Deep Reinforcement Learning}
    \acro{RL}{Reinforcement Learning}
    \acro{DNN}{Deep Neural Network}
    \acro{LSTM}{Long Short-Term Memory}
    \acro{CNN}{Convolutional Neural Network}
    \acro{D3QL}{Dueling Double Deep Q-learning}

    \acro{SLA}{Service-Level Agreement}
    \acro{6G}{Sixth Generation}
    \acro{MEC}{Mobile Edge Computing}

    \acro{RIS}{Reconfigurable Intelligent Surface}
    \acro{NTN}{Non-Terrestrial Network}

    \acro{Mbps}{Megabit Per Second}
    \acro{Gbps}{Gigabit Per Second}
    \acro{Tbps}{Terabit Per Second}
    \acro{GFLOPS}{Giga Floating-Point Operations Per Second}
    \acro{SUMO}{Simulation of Urban Mobility}

    \acro{DAG}{Directed Acyclic Graph}
    \acro{HPC}{High Performance Computing}
    \acro{SLO}{Service Level Objective}

    \acro{HRLLC}{Hyper Reliable and Low-Latency Communication}
    \acro{eMBB}{Enhanced Mobile Broadband}
    \acro{mMTC}{Massive Machine-Type Communication}
    \acro{NOMA}{Non Orthogonal Multiple Access}
    \acro{PPO}{Proximal Policy Optimization}
    \acro{MAPPO}{Multi-Agent Proximal Policy Optimization}
    \acro{MMDP}{Markov Decision Process}
    \acro{GAE}{Generalized Advantage Estimation}
\end{acronym}

\title{Multi-Agent Reinforcement Learning for SLA-Aware Network Slicing in UAV-Enabled MEC}

\author{
    \IEEEauthorblockN{
        Mohammad Farhoudi\textsuperscript{1}, Zeinab Sasan\textsuperscript{2}, Masoud Shokrnezhad\textsuperscript{3}, and Tarik Taleb\textsuperscript{4}\\
    }
    \IEEEauthorblockA{
        \textsuperscript{1} \textit{Oulu University, Oulu, Finland}; mohammad.farhoudi@oulu.fi \\
        \textsuperscript{2} \textit{Amirkabir University of Technology, Tehran, Iran}; z.sasan@aut.ac.ir \\
        \textsuperscript{3} \textit{ICTFICIAL Oy, Espoo, Finland}; masoud.shokrnezhad@ictficial.com \\
        \textsuperscript{4} \textit{Ruhr University Bochum (RUB), Bochum, Germany}; tarik.taleb@rub.de \\ 
    }
}

\maketitle

\begin{abstract}
    Unmanned Aerial Vehicle (UAV)-enabled Mobile Edge Computing (MEC) offers flexible capacity provisioning for heterogeneous network slices, including Hyper-Reliable and Low-Latency Communication (HRLLC), Enhanced Mobile Broadband (eMBB), and Massive Machine-Type Communications (mMTC). However, guaranteeing slice-level Service-Level Agreements (SLAs) under dynamic user mobility, stochastic task arrivals, and constrained onboard energy and computing resources remains a fundamental challenge. This paper proposes a predictive multi-agent Reinforcement Learning (RL) framework that proactively maintains SLA stability in UAV-enabled MEC through coordinated trajectory control and computation resource allocation. A lightweight prediction module forecasts near-future user mobility, enabling UAVs to anticipate congestion and reposition before SLA violations occur. We design an SLA-aware reward function that explicitly penalizes both violation probability and duration across slices, alongside total energy consumption. UAV agents are trained using Multi-Agent Proximal Policy Optimization (MAPPO) with centralized training and decentralized execution, enabling scalable online decision-making. Event-driven simulations with realistic mobility traces demonstrate that the proposed framework significantly improves SLA stability compared with baselines while maintaining competitive energy efficiency and delay performance, approaching oracle-level performance with sufficiently accurate predictive information.
\end{abstract}


\section{Introduction}

The rapid proliferation of computation-intensive and delay-sensitive applications, such as augmented reality, autonomous systems, and real-time video analytics, has imposed stringent requirements on next-generation wireless networks \cite{farhoudi_discovery_2025}. \Ac{MEC} has emerged as a key enabler to address these challenges by bringing computational resources closer to end users, thereby reducing delay and alleviating backhaul congestion \cite{10855598}. Meanwhile, \acp{UAV}, due to their flexibility, rapid deployment, and communication capabilities, have been increasingly integrated into \ac{MEC} systems to provide on-demand edge services in scenarios with limited or damaged infrastructure, such as remote monitoring and temporary hotspots. In parallel, network slicing by logically partitioning network resources into multiple isolated slices, enables customized service provisioning for applications with distinct performance requirements, such as \ac{HRLLC}, \ac{eMBB}, and \ac{mMTC} \cite{sasan2025balancing}, \cite{sasan2024joint}. The integration of \ac{UAV}-enabled \ac{MEC} with network slicing offers a promising paradigm for delivering flexible and efficient edge intelligence in dynamic environments.


However, realizing this vision introduces significant technical challenges. In \ac{UAV}-enabled \ac{MEC} systems with network slicing, multiple \acp{UAV} should serve ground users with heterogeneous slice requirements while jointly optimizing their trajectory planning, user association, and computation resource allocation. Each slice imposes distinct \ac{SLA} constraints on tolerable delay, requiring careful coordination between communication and computation resources. The problem is further complicated by \ac{UAV} mobility constraints, limited onboard energy budgets, and constrained computation capacity. Moreover, the system should operate under dynamic and uncertain conditions, including time-varying user mobility, stochastic task arrivals with different characteristics, and evolving channel conditions. These factors result in a complex, stochastic, and time-coupled optimization problem where current \ac{UAV} positions and energy states influence future system dynamics. 


\begin{table*}[t]
    \centering
    \caption{Comparison of Existing Works and the Proposed Method}
    \label{tab:comparison}
    \vspace{-6pt}
    \setlength{\tabcolsep}{4pt}
    \resizebox{\textwidth}{!}{%
        \begin{tabular}{c p{6.6cm} c c c c c c p{5.5cm}}
            \toprule
            \textbf{Ref.}
             & \textbf{Main Focus}
             & \textbf{UAV-MEC}
             & \textbf{Slicing}
             & \textbf{Trajectory}
             & \textbf{Learning}
             & \textbf{SLA-Aware}
             & \textbf{Predictive}
             & \textbf{Main Limitation}                                                                                              \\
            \midrule

            \cite{farhoudi2025deep}
             & Service composition in aerial-terrestrial networks
             & \cmark                                                                   & \xmark & \xmark & \cmark & \xmark & \cmark
             & No slicing and no SLA-aware control                                                                                   \\

            \cite{farhoudi2026energy}
             & Energy-efficient orchestration in 6G aerial-terrestrial
             & \cmark                                                                   & \xmark & \cmark & \cmark & \xmark & \cmark
             & Focus on energy and QoS, no slicing                                                                                   \\

            \cite{wu2023intelligent}
             & Survivable resource slicing in UAV-MEC
             & \cmark                                                                   & \cmark & \xmark & \cmark & \xmark & \cmark
             & No trajectory/offloading joint optimization                                                                           \\

            \cite{tang2022slicing}
             & SDN-based slicing architecture for UAV-MEC
             & \cmark                                                                   & \cmark & \xmark & \xmark & \xmark & \xmark
             & Mostly architectural, no dynamic control                                                                              \\

            \cite{faraci2020design}
             & 5G slice extension with UAV-MEC
             & \cmark                                                                   & \cmark & \xmark & \cmark & \xmark & \xmark
             & Only slice extension, no multi-slice orchestration                                                                    \\

            \cite{tian2023service}
             & User satisfaction-based task offloading
             & \cmark                                                                   & \xmark & \cmark & \xmark & \xmark & \xmark
             & No slicing and no SLA guarantees                                                                                      \\

            \cite{chen2025qos}
             & QoS-aware task offloading in multi-UAV MEC
             & \cmark                                                                   & \xmark & \cmark & \cmark & \xmark & \xmark
             & QoS-based, no SLA modeling                                                                                            \\

            \cite{li2025self}
             & Dynamic self-adjusting network slicing
             & \cmark                                                                   & \cmark & \cmark & \cmark & \xmark & \cmark
             & No explicit SLA violation modeling                                                                                    \\

            \midrule
            \textbf{This work}
             & \textbf{SLA-stable slicing with predictive multi-agent learning}
             & \cmark                                                                   & \cmark & \cmark & \cmark & \cmark & \cmark
             & \textbf{SLA-aware slicing, trajectory, and offloading}                                                          \\
            \bottomrule
        \end{tabular}%
    }
    \vspace{-6pt}
\end{table*}

Extensive research has investigated \ac{UAV}-enabled \ac{MEC} systems from multiple perspectives. Several works applied deep \ac{RL} approaches to optimize \ac{UAV} trajectory and resource orchestration in \ac{UAV}-enabled systems, focusing primarily on system efficiency and energy consumption \cite{farhoudi2025deep,farhoudi2026energy}. In parallel, some studies introduced network slicing concepts into \ac{UAV}-\ac{MEC} systems. For instance, \cite{wu2023intelligent,tang2022slicing,faraci2020design} proposed slicing frameworks that emphasize resource partitioning, survivability mechanisms, and architectural design for heterogeneous service provisioning. Another research direction addressed task offloading optimization, where Tian~\textit{et al.} \cite{tian2023service} and Chen~\textit{et al.} \cite{chen2025qos} developed user satisfaction and \ac{QoS}-oriented offloading schemes in multi-\ac{UAV} settings. Also, Li \textit{et al.} \cite{li2025self} introduced a self-adjusting network slicing mechanism using two-timescale \ac{RL}, which adapts slice configurations based on network dynamics, representing an advancement toward integrating slicing and learning-based control in \ac{UAV}-\ac{MEC} systems.

Despite these advancements, existing works mainly optimize conventional \ac{QoS} metrics such as delay or throughput. However, such metrics are insufficient for guaranteeing \acp{SLA}, which require strict and often probabilistic guarantees on performance metrics. As shown in Table~\ref{tab:comparison}, most existing approaches treat slicing, trajectory control, and resource orchestration as separate problems, or rely on reactive mechanisms that adapt only after performance degradation occurs. In practical multi-service environments, different slices have heterogeneous and time-varying requirements, and maintaining stable SLA satisfaction under user mobility, stochastic traffic arrivals, and \ac{UAV} energy constraints remains a critical challenge. As a result, the fundamental problem of \emph{SLA stability in \ac{UAV}-enabled MEC network slicing} remains largely unexplored.

To address these challenges, this paper proposes a novel framework for \emph{SLA-aware network slicing in \ac{UAV}-enabled MEC systems}, where \acp{UAV} serve as dynamic orchestrators for maintaining \ac{SLA} guarantees across multiple service slices. The main paper's contributions are summarized as follows:


\begin{itemize}
    \item We formulate a joint optimization problem that integrates \ac{UAV} trajectory control, user association, and slice-level resource allocation to minimize energy consumption and \ac{SLA} violation probability and duration across slices.
    \item We develop a predictive multi-agent \ac{RL} framework based on \ac{MAPPO}, where each \ac{UAV} leverages user mobility predictions to proactively prevent \ac{SLA} violations.
    \item Simulations demonstrate that the proposed approach outperforms baseline methods in terms of SLA satisfaction, temporal stability, and energy efficiency.
\end{itemize}

In the rest: Section~\ref{sec:formulation} presents the system model and problem formulation, Section~\ref{sec:method} describes the proposed methodology, Section~\ref{sec:results} evaluates the performance through simulations, and Section~\ref{sec:conclusion} concludes the paper.
\section{System Model and Problem Formulation}\label{sec:formulation}

\begin{figure}[t!]\centering
\vspace{-2pt}
\includegraphics[width=3.5in]{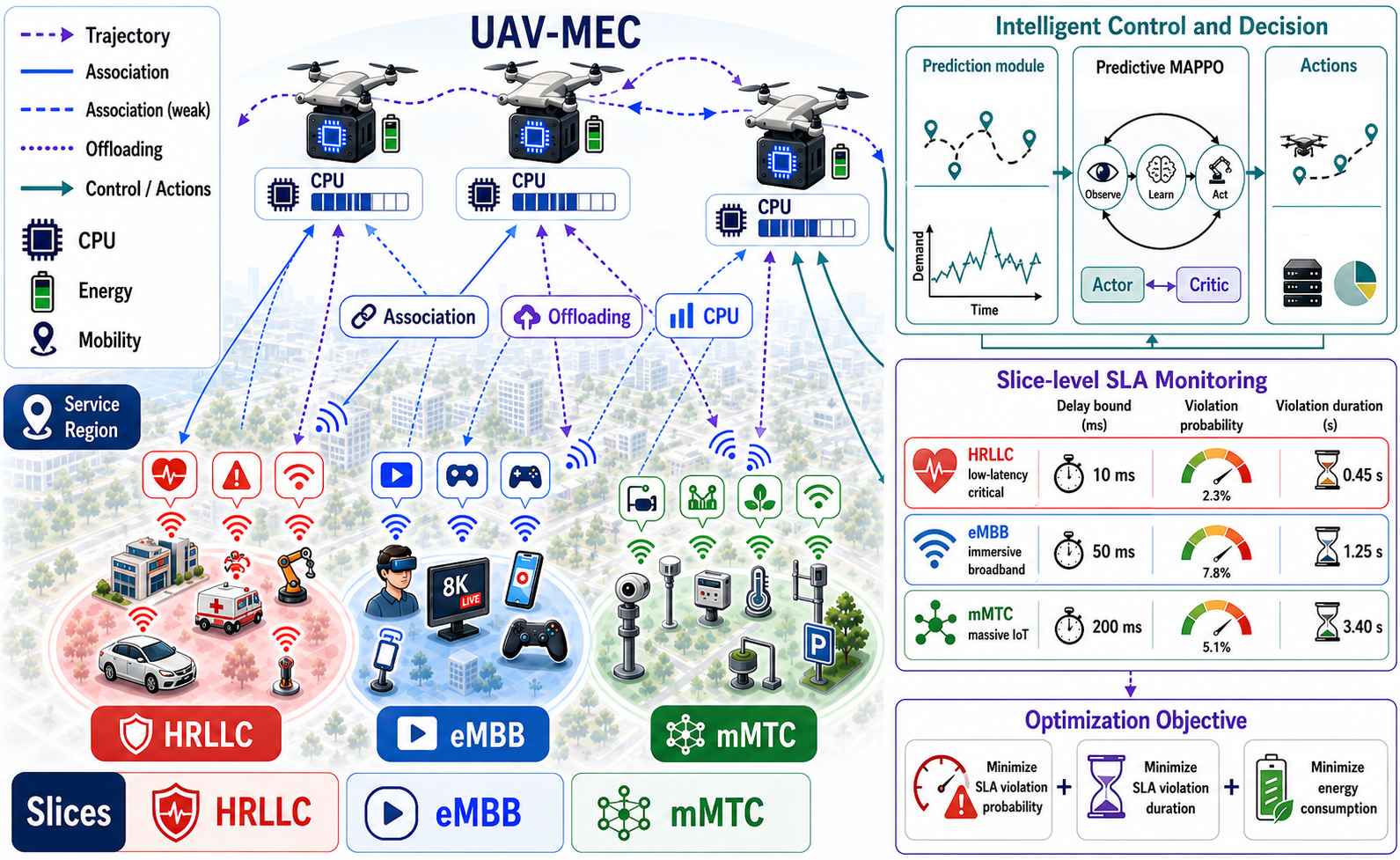}
\vspace{-17pt}
  \caption{System model, including ground users grouped into heterogeneous slices with slice-level SLA requirements. UAVs act as flying edge servers.}
    \label{figure:system_model}
    \vspace{-4pt}
\end{figure}


We consider a \ac{UAV}-enabled \ac{MEC} system, where \acp{UAV} provide computation offloading services for ground users, as depicted in Fig.~\ref{figure:system_model}. The system operates over a finite time horizon $\boldsymbol{\mathcal{T}}$, divided into discrete time slots indexed by $t$ and with duration $\Delta t$. The set of ground users is denoted by $\boldsymbol{\mathcal{K}}\!~=~\!\{1,\dots,K\}$. Users generate computation-intensive tasks, which are processed within the same time slot in which they are generated, and offloaded to \acp{UAV} for remote execution. To capture dynamic task arrivals, we define a binary task-arrival indicator $\lambda_k(t)\in\{0,1\}$, where $\lambda_k(t)=1$ indicates that user $k$ generates a computation task at time slot $t$. The set of \acp{UAV} is denoted by $\boldsymbol{\mathcal{U}}=\{1,\dots,U\}$, where each is equipped with communication and computation capabilities, acting as flying edge servers. The maximum computation capacity $F_u^{\max}$, maximum speed $V^{\max}_u$, and energy budget $E_u^{\max}$ for each \ac{UAV} $u$ represent constrained onboard resources.

\subsubsection{Network Slicing Model}

The system supports multiple network slices to serve heterogeneous applications with diverse service requirements. The set of slices is denoted by $\boldsymbol{\mathcal{S}}=\{1,\dots,S\}$, where each slice $s$ represents a logical service class such as \ac{HRLLC}, \ac{eMBB}, or \ac{mMTC}, aligned with representative \ac{6G} usage scenarios \cite{ITUR-M2160}.
The class of slice $s$ specifies its offload task's profile, defined as 
\begin{equation}
\mathcal{\xi}_s=\{\bar{D}_s,\bar{C}_s,\bar{\tau}_s^{\max}\},
\end{equation}
where $\bar{D}_s$ represents the nominal input data size, $\bar{C}_s$ denotes the nominal required CPU cycles, and $\tau_s^{\max}$ is the maximum tolerable delay for slice $s$. We denote the set of active users associated with slice $s$ by $\mathcal{K}_s(t)$, where $\mathcal{K}_s(t) \subseteq \boldsymbol{\mathcal{K}}$.



The generated task of user $k$ belongs to the slice $s_k$, whose characteristics are determined by the corresponding profile $\mathcal{\xi}_{s_k}$. In particular, $\tau_k^{\max}\!\!=\!\!\bar{\tau}_{s_k}^{\max}$, while $D_k(t)$ and $C_k(t)$ are generated according to slice-dependent distributions around $\bar{D}_{s_k}$ and $\bar{C}_{s_k}$. The task is expressed by $\{D_k(t), C_k(t), \tau_k^{\max}, \mathbf{w}_k(t)\}$ with $\mathbf{w}_k(t)=[x_k(t),y_k(t),z_k(t)]$ represent the location of $k$ at time slot $t$, where $x_k(t)$ and $y_k(t)$ refer to the horizontal coordinates, and $z_k(t)\!\!=\!0$ denotes the altitude for ground users.

\subsubsection{UAV Mobility Model}

\ac{UAV} $u$ moves in a three-dimensional space with position $\mathbf{q}_u(t)=[x_u(t),y_u(t),z_u(t)]$ that directly affects the distance to users. Due to physical mobility limitations, the displacement of each \ac{UAV} between two consecutive time slots is constrained by $V^{\max}_u$. Accordingly, the mobility constraint of \ac{UAV} $u$ is expressed as
\begin{equation}
    \label{eq:uav_mobility}
    \mathsmaller{\|\mathbf{q}_u(t+1)-\mathbf{q}_u(t)\|\leq V^{\max}_u\Delta t, \quad \forall u\in\boldsymbol{\mathcal{U}},\, t\in\boldsymbol{\mathcal{T}}}.
\end{equation}

\subsubsection{User Association Model}
Active user $k$ is associated with and offloads its task to \ac{UAV} $u$, which is indicated by a binary variable $a_{k,u}(t) \in \{0,1\}$.
Specifically, $a_{k,u}(t)=1$ means that user $k$ offloads its task to \ac{UAV} $u$ ($0$ otherwise). Each active user is assumed to be served by one \ac{UAV} at each time slot; therefore, the association constraint is expressed as
\begin{equation}
    \label{eq:association}
    \mathsmaller{\sum_{u\in\boldsymbol{\mathcal{U}}} a_{k,u}(t)=\lambda_k(t),\quad \forall k\in\boldsymbol{\mathcal{K}}, t\in\boldsymbol{\mathcal{T}}}.
\end{equation}

We assume that users maintain connectivity with the selected \ac{UAV}, where larger distances are reflected through reduced transmission rates.

\subsubsection{Communication Model}

The achievable transmission rate between user $k$ and \ac{UAV} $u$ depends on (i) their relative distance $d_{k,u}(t)\!=\!\left\| \mathbf{q}_u(t) - \mathbf{w}_k(t) \right\|$, (ii) channel conditions, and (iii) transmit power.
The channel gain is modeled using a distance-dependent path-loss exponent $\gamma$ and channel gain $h_{k,u}(t)\!=\!\frac{\beta_0}{d_{k,u}^{\gamma}(t)}$ at a reference distance $\beta_0$. The transmit power is not treated explicitly as an optimization variable in practical modeling; rather, it is taken as a distance-aware power control mechanism, where users adapt their transmit power based on the communication distance. Specifically, the transmit power of user $k$ when communicating with \ac{UAV} $u$ is modeled as
\begin{equation}
    \mathsmaller{P_{k,u}(t) = P_0 \left(\frac{d_{k,u}(t)}{d_0}\right)^{\rho},}
\end{equation}
where $P_0$ is the reference transmit power at distance $d_0$, and $\rho$ controls the degree of path-loss compensation. In particular, $\rho < \gamma$ indicates partial compensation, in which the received signal quality degrades with distance. With channel bandwidth $B$ and noise power $\sigma^2$, the transmission rate is given by
\begin{equation}
    \mathsmaller{R_{k,u}(t)=B\log_2\left(1+\frac{P_{k,u}(t) \, h_{k,u}(t)}{\sigma^2}\right)}.
\end{equation}
Accordingly, the transmission delay required to upload the input data to the selected \ac{UAV} is given by
\begin{equation}
    \mathsmaller{T_k^{\mathrm{tx}}(t)=\sum_{u\in\boldsymbol{\mathcal{U}}} a_{k,u}(t)\frac{D_k(t)}{R_{k,u}(t)}}.
\end{equation}

\subsubsection{Computation Model}

Each \ac{UAV} allocates its CPU resource to its associated users. The amount of CPU cycles allocated by \ac{UAV} $u$ to user $k$ at time slot $t$ is denoted by $f_{k,u}(t)$, which should satisfy $f_{k,u}(t) \geq 0$.
The allocated resources should be sufficient to complete the task within the considered time scale. The computation model is defined as
\begin{equation}
    \mathsmaller{T_k^{\mathrm{comp}}(t)=\sum_{u\in\boldsymbol{\mathcal{U}}} a_{k,u}(t)\frac{C_k(t)}{f_{k,u}(t)}}.
\end{equation}

To model the resource limitations, the total allocated resources $f_{k,u}(t)$ for all connected users cannot exceed UAV $u$'s maximum capacity, expressed as
\begin{equation}
    \label{eq:max_capacity}
    \mathsmaller{\sum_{k\in\boldsymbol{\mathcal{K}}} a_{k,u}(t) f_{k,u}(t)\leq F_u^{\max},\quad \forall u\in\boldsymbol{\mathcal{U}}, t\in\boldsymbol{\mathcal{T}}}.
\end{equation}


\subsubsection{SLA Violation}

The SLA violation manifests itself in three ways: (i) user-level SLA violation, (ii) slice-level SLA violation, and (iii) SLA violation duration. User-level violation occurs when the task completion delay exceeds the task's maximum tolerable delay, expressed as 
\begin{equation}
    \mathsmaller{I_k(t)=1 \text{ if } \tau_k(t)=T_k^{\mathrm{tx}}(t)+T_k^{\mathrm{comp}}(t)> \tau_k^{\max}}.
\end{equation}
At each time slot $t$, the instantaneous SLA violation ratio of slice $s$ is defined as
\begin{equation}
    \mathsmaller{P_s^{\mathrm{viol}}(t) =
    \frac{1}{|\mathcal{K}_s(t)|}
    \sum_{k\in\mathcal{K}_s(t)} I_k(t),}
\end{equation}
which denotes the fraction of users in $\mathcal{K}_s(t)$ whose \acp{SLA} are violated. We introduced the long-term SLA violation as $\bar{P}_s^{\mathrm{viol}}=\frac{1}{|\boldsymbol{\mathcal{T}}|} \sum_{t\in\boldsymbol{\mathcal{T}}} P_s^{\mathrm{viol}}(t)$ that captures the SLA violation experienced by $s$ over time. 
Since SLA degradation may persist over multiple time slots, we also quantify its temporal persistence by defining the normalized SLA violation duration of slice $s$ over the time horizon as
\begin{equation}
    \mathsmaller{\tilde{T}_s^{\mathrm{viol}}=
    \frac{1}{|\boldsymbol{\mathcal{T}}|}
    \sum_{t\in\boldsymbol{\mathcal{T}}}
    \mathbb{I}\left(P_s^{\mathrm{viol}}(t)>\epsilon_s\right),}
\end{equation}
where $\epsilon_s$ is a predefined violation threshold, and $\mathbb{I}(\cdot)$ denotes the indicator function. 

\subsubsection{UAV Energy Model}

The total \ac{UAV} energy consumption over the time horizon consists of (i) propulsion (flight) energy for movement and (ii) computation energy for task processing
\begin{equation}
\label{eq:total_energy}
    \mathsmaller{E_u = \sum_{t\in\boldsymbol{\mathcal{T}}} \left( E_u^{\mathrm{fly}}(t) + E_u^{\mathrm{comp}}(t) \right) \leq E_u^{\max}, \; \forall u \in \boldsymbol{\mathcal{U}},}
\end{equation}
that should not exceed the available energy budget. The propulsion energy is modeled as a tractable approximation of \ac{UAV} displacement between consecutive time slots
\begin{equation}
    \mathsmaller{E_u^{\mathrm{fly}}(t) = \varsigma \left\| \mathbf{q}_u(t+1) - \mathbf{q}_u(t) \right\|^2,}
\end{equation}
where $\varsigma$ is a propulsion-energy coefficient. With the energy consumption coefficient per CPU cycle $\eta$, the computation energy utilized at time slot $t$ is modeled as
\begin{equation}
    \mathsmaller{E_u^{\mathrm{comp}}(t) = \eta \sum_{k\in\boldsymbol{\mathcal{K}}} a_{k,u}(t) f_{k,u}(t)}. 
\end{equation}


\subsubsection{Problem Formulation}

The optimization objective
\begin{equation}
    \min_{\mathbf{q},a,f}
     \sum_{s\in\boldsymbol{\mathcal{S}}} \! b_s \bar{P}_s^{\mathrm{viol}}
    + \sum_{s\in\boldsymbol{\mathcal{S}}}  \beta_s \tilde{T}_s^{\mathrm{viol}}
    + \chi \! \sum_{u\in\boldsymbol{\mathcal{U}}} \! E_u \quad \text{s.t.} \, \eqref{eq:uav_mobility}, \eqref{eq:association}, \eqref{eq:max_capacity}, \eqref{eq:total_energy} \notag
\end{equation}
jointly optimizes \ac{UAV} trajectory, user association, and resource allocation to improve SLA stability while controlling energy consumption. The coefficients $b_s$, $\beta_s$, and $\chi$ are weighting parameters that control the SLA violation probability, SLA violation duration, and UAV energy consumption trade-off. Decision variables govern \ac{UAV} trajectory $\mathbf{q}_u(t)$, user association $a_{k,u}(t)$, and computation resource allocation $f_{k,u}(t)$. 

The problem is non-convex, stochastic, and time-coupled due to binary decisions, nonlinear rates, dynamic task arrivals, user mobility, and evolving UAV energy states. These challenges limit real-time optimal solutions and motivate efficient suboptimal and learning-based approaches. Also, purely reactive strategies, which rely only on current system observations, are insufficient in highly dynamic environments, motivating the need for predicting future tasks and mobility patterns for maintaining SLA satisfaction.

\section{Proposed Predictive Multi-Agent Framework}\label{sec:method}
\begin{figure}[t!]\centering
    \vspace{-4pt}
    \includegraphics[width=3.5in]{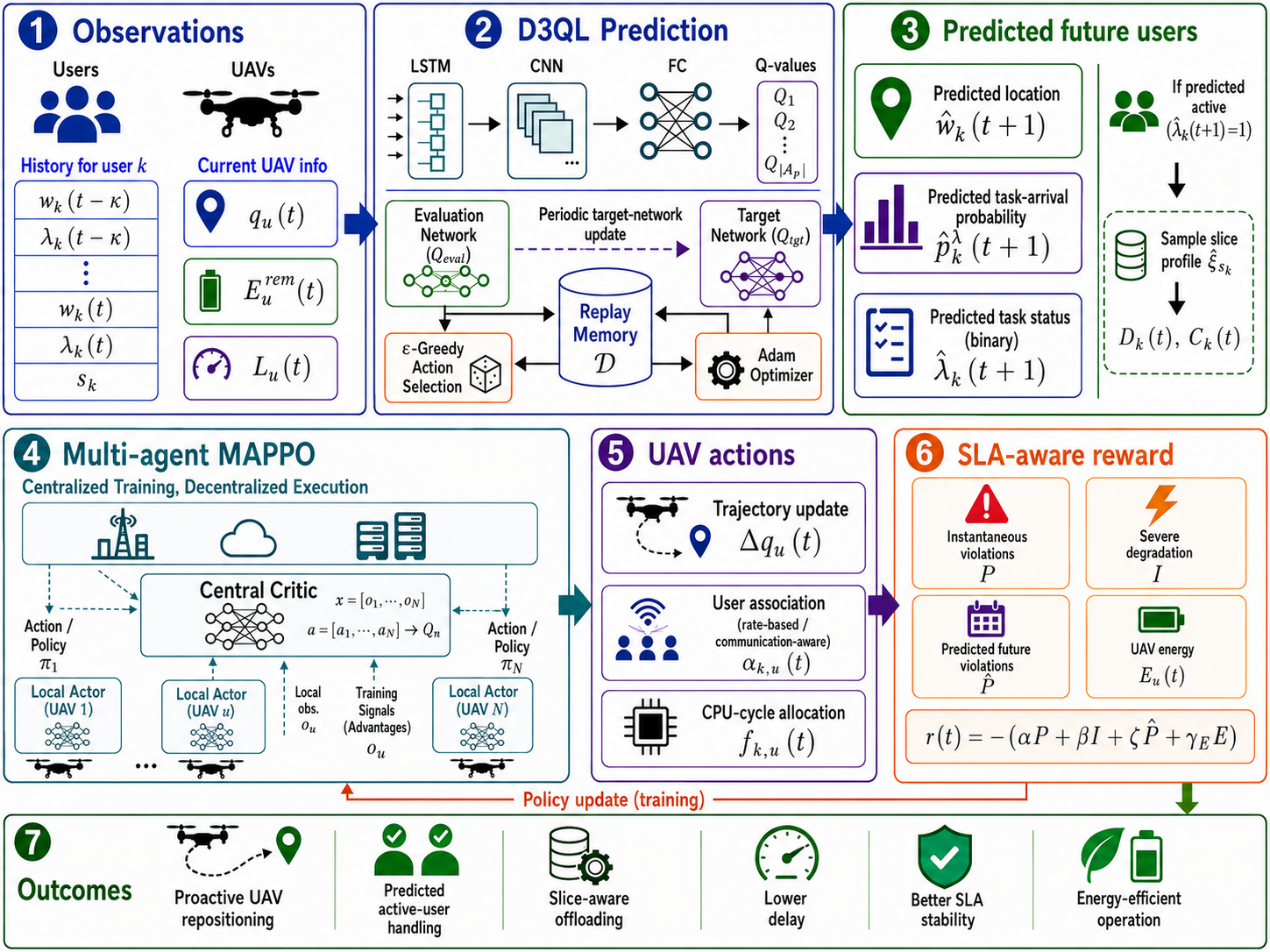}
    \vspace{-18pt}
    \caption{Proposed framework: D3QL-based user prediction with MAPPO-based UAV trajectory control, rate-based association, and computation allocation.}
    \label{figure:method}
    \vspace{-8pt}
\end{figure}

To enable proactive control in \ac{UAV}-enabled \ac{MEC} systems, we propose a \ac{SLA}-aware predictive multi-agent network slicing framework, depicted in Fig.~\ref{figure:method}. The proposed framework consists of three main components: (i) a prediction module that estimates user locations and task-generation status, (ii) a decentralized multi-agent decision-making module in which each \ac{UAV} acts as an autonomous agent, and (iii) a centralized \ac{SLA}-aware policy optimization mechanism that jointly penalizes instantaneous \ac{SLA} violations, persistent slice-level degradation, predicted violations, and \ac{UAV} energy consumption. The main idea is to incorporate the predictions into the decision-making process that enables \acp{UAV} to anticipate future communication and computation pressures and adjust their trajectories, user association, and computation resource allocation.

\subsection{Prediction}

This module estimates the near-future location $\hat{\mathbf{w}}_k(t+1)$ and task-generation probability $\hat{p}^{\lambda}_k(t+1)$ for each user $k$. It utilizes a deep \ac{RL} algorithm with model $\mathcal{F}_{\psi}(\cdot)$, parameterized by $\psi$, and formulates the prediction process as a Markov decision process over the mobility-region action space. The prediction state, constructed from $\kappa$ observation windows, is defined as
\begin{equation}
    \mathsmaller{
        \Omega_k(t)=
        \left\{
        \mathbf w_k(t-\kappa),\lambda_k(t-\kappa),
        \ldots,
        \mathbf w_k(t),\lambda_k(t),s_k
        \right\}}.
\end{equation}
The prediction action $\alpha_k(t)$ is $(\hat{\mathbf{w}}_k(t+1), \hat{p}^{\lambda}_k(t+1))$. 
After the actual user location and task-generation status at time slot $t+1$ are observed, the prediction reward is computed as
\begin{equation}
    \!\!\!\rho_k(t)\!=\!\omega_{\mathrm{\alpha}}\mathbb{I}(\hat{\alpha}_k(t\!+\!1)\!\!=\!\!\alpha_k(t\!+\!1))+\omega_{\lambda}\mathbb{I}(\hat{\lambda}_k(t\!+\!1)\!\!=\!\!\lambda_k(t\!+\!1)),
\end{equation}
where $\omega_{\alpha}$ and $\omega_{\lambda}$ weight mobility versus task-generation accuracy. For predicted active users, the data size $D_k(t)$ and CPU cycles $C_k(t)$ are sampled according to the slice-dependent distributions defined by the slice profile $\xi_{s_k}$.

We employ a \ac{D3QL} architecture for the prediction model, as it improves the stability of Q-value estimation by combining double Q-learning and dueling network decomposition. The predictor's neural structure follows a hybrid recurrent-convolutional design: (i) the historical sequence is first processed by an \ac{LSTM} layer to capture temporal dependencies in movement and generation behavior; (ii) then passed through convolutional layers to extract local transition patterns from the encoded sequence; and (iii) fully connected layers map the extracted features to Q-values over the prediction action space $\mathcal A_p$. The prediction action is selected according to an exploration-exploitation policy that selects the action with the highest Q-value $\alpha_k(t)=\arg\max_{\alpha\in\mathcal{A}_p} Q(\Omega_k(t),\alpha;\psi)$, or selects a random action to encourage exploration.

\vspace{-2pt}
\subsection{Decentralized Multi-agent Decision-Making}
The predicted information is incorporated into a multi-agent Markov decision process, where each UAV acts as an agent. We adopt centralized training with decentralized execution: a centralized critic uses the global state to evaluate the joint behavior of UAV agents, while each UAV independently selects its action using only its local observation and policy $\!\pi_{\theta_u}\!$. At time slot $t$, the local observation of UAV $u$ is
\begin{align}
o_u(t)=
\Big\{
& q_u(t),\ E_u^{re}(t),\ L_u(t), \big\{
s_k,\ D_k(t),\ C_k(t),\ w_k(t), \notag \\
& \quad \hat{w}_k(t+1),\ \hat{p}^{\lambda}_k(t+1),\ \tau_k^{\max}
\big\}_{k \in \mathcal{K}^{obs}_u(t)}
\Big\}.
\end{align}
Here, $E_u^{re}(t)$ represents its remaining energy, $L_u(t)$ denotes its current computational load, and $\mathcal{K}^{obs}_u(t)$ shows the users that are predicted to be active and observable by $u$. 
Additionally, $\hat{w}_k(t+1)$ and $\hat{p}^{\lambda}_k(t+1)$ denote the predicted next-slot location and task-generation probability, respectively.



Each UAV agent selects an action that controls its movement and the computation resource allocation vector for the users in its observation set:
\begin{align}
A_u(t)=\left\{\Delta q_u(t), \mathbf{f}_u(t)\right\},
\;
\mathbf{f}_u(t)=
\left\{
f_{k,u}(t)
\right\}_{k \in \mathcal{K}^{obs}_u(t)},
\end{align}
where $\Delta q_u(t)$ denotes the displacement of UAV $u$ at time $t$, such that
$q_u(t+1)=q_u(t)+\Delta q_u(t)$ while satisfying Eq.~\eqref{eq:uav_mobility}. 
The vector $\mathbf{f}_u(t)$ denotes the CPU-allocation vector of UAV $u$ over its observed user set $\mathcal{K}^{obs}_u(t)$, and each element $f_{k,u}(t)$ specifies the computation resource assigned to user $k$. 
The allocated resources are constrained by the maximum computation capacity of each UAV, as given in Eq.~\eqref{eq:max_capacity}. 
Given trajectories and channels at $t$, user association follows $a_{k,u}(t)\!=\!1 \, \text{if } u=\arg\!\max\limits_{j\in\boldsymbol{\mathcal{U}}}R_{k,j}(t)$ (and $0$ otherwise), i.e., each user attaches to the UAV with the highest achievable rate.

\subsection{Centralized SLA-aware Policy Optimization}
We adopt \ac{MAPPO}~\cite{yu2022MAPPO} to optimize the policies of UAV agents, as it supports cooperative multi-agent learning under centralized training and decentralized execution. After taking actions independently, the agents cooperate through the shared reward signal $r(t)$ designed to encourage SLA-aware and energy-efficient behavior, defined as
\begin{align}
    \mathsmaller{r(t)=-\Big(}
     & \mathsmaller{\sum_{s\in\boldsymbol{\mathcal{S}}}b_s P_s^{\mathrm{viol}}(t)
    +\sum_{s\in\boldsymbol{\mathcal{S}}}\beta_s
    \mathbb{I}\left(P_s^{\mathrm{viol}}(t)>\epsilon_s\right) \nonumber}+               \\
     & \mathsmaller{\sum_{s\in\boldsymbol{\mathcal{S}}}b_s
    \hat{P}_s^{\mathrm{viol}}(t+1)
    +\chi\sum_{u\in\boldsymbol{\mathcal{U}}}E_u(t)
    \Big)}.
\end{align}
In shared reward, $\hat{P}_s^{\mathrm{viol}}(t{+}1)$ is the slice violation ratio computed from predicted positions $\hat{\mathbf{w}}_k(t{+}1)$, predicted activity $\hat{\lambda}_k(t{+}1)$, max-rate association under predicted geometry at $t{+}1$, and sampled $(\hat{D}_k,\hat{C}_k)$ when active; use $0$ if no user is predicted active in slice $s$. The reward penalizes (i) instantaneous slice-level SLA violations, (ii) severe degradation to avoid persistent violating states, (iii) predicted near-future violations estimated from $\hat{\mathbf{w}}_k(t+1)$, and (iv) UAV energy consumption $E_u(t)$ with $b_s$, $\beta_s$, and $\chi$ controlling the trade-off between SLA satisfaction and energy efficiency.

The objective of policy optimization is to maximize the expected discounted cumulative reward $\mathbb{E}\left[\sum_{t=0}^{\infty}\mu_{\mathrm{RL}}^t r(t) \right],$
where $\mu_{\mathrm{RL}}\in(0,1)$ is the \ac{RL} discount factor. Shared rewards enter the policy update through the temporal-difference error
\begin{equation}
    \mathsmaller{\delta(t)=r(t)+\mu_{\mathrm{RL}}V_{\phi}(\mathbf{s}(t+1))-V_{\phi}(\mathbf{s}(t)),}
\end{equation}
where $\!V_{\phi}(\cdot)\!$ is the centralized critic and $\mathbf{s}(t)=\{o_u(t)\}_{u\in\boldsymbol{\mathcal{U}}}$ is the joint state. The advantage function is then computed via \ac{GAE}, given by
\begin{equation}
    \mathsmaller{\hat{A}(t)=
        \sum_{l=0}^{\infty}
        (\mu_{\mathrm{RL}}\lambda_{\mathrm{GAE}})^l
        \delta(t+l),}
\end{equation}
where $\lambda_{\mathrm{GAE}}$ is the \ac{GAE} smoothing parameter. Each UAV policy is then updated using the PPO clipped surrogate objective
\begin{equation}
    \!\!\!\!\!\mathsmaller{L^{\mathrm{O}}\!(\theta_u)
    \!\!=\!\!
    \mathbb{E}(t)
    [
    \min
    (
    \rho_{\theta_u}\!(t)\!\hat{A}(t),
    \mathrm{clp}(\rho_{\theta_u}\!\!(t),1\!-\!\epsilon,1\!\!+\!\epsilon)\hat{A}(t)
    )
    ]},\!\!\!
\end{equation}
where $\epsilon$ is the PPO clipping coefficient used to prevent excessively large policy updates and stabilize learning. Finally, the probability ratio for UAV agent $u$ is defined by
\begin{equation}
    \mathsmaller{\rho_{\theta_u}(t)=
    \frac{
    \pi_{\theta_u}(A_u(t)|o_u(t))
    }{
        \pi_{\theta_u^{\mathrm{old}}}(A_u(t)|o_u(t))
        }}.
\end{equation}
\section{Performance Evaluation}\label{sec:results}
We evaluate the proposed \emph{predictive multi-agent slicing} framework via event-driven simulations in terms of total UAV energy consumption (propulsion + computation), average service delay, and the SLA stability metrics defined in Section~\ref{sec:formulation}. 
Unless otherwise stated, we simulate $U{=}3$ UAVs serving $K{=}24$ users in a $1000{\times}1000$~m$^2$ area with heterogeneous \ac{HRLLC}/\ac{eMBB}/\ac{mMTC} task profiles and slice-specific thresholds (Table~\ref{tab:simulation_parameters}). Users follow the YJMob100K mobility traces~\cite{yabe2024yjmob100k}, while task sizes and CPU cycles are sampled around the profiles, consistent with Section~\ref{sec:formulation}. The proposed method leverages the predictor in Section~\ref{sec:method} to incorporate $\hat{\mathbf{w}}_k(t{+}1)$ and predicted activity into the \ac{MAPPO}; in all experiments, we report averages over the evaluation episodes. To perform the analysis, we conduct two scenarios.

\begin{table}[t]
    \caption{Simulation Parameters}
    \vspace{-6pt}
    \label{tab:simulation_parameters}
    \centering
    \footnotesize
    \begin{tabular}{ll}
        \toprule
        \textbf{Parameter}                                                & \textbf{Value}                                                    \\
        \midrule
        Area / UAVs / users                                               & $1000 \times 1000$ m$^2$ / $3$ / $24$                             \\
        UAV altitude/speed/CPU/energy                                     & $80$-$140$m / $30$mps /$12$GHz / $5\!\times\!10^7$J               \\
        Bandwidth / noise / Path-loss                                     & $8$ MHz / $10^{-13}$ W / $2.1$                                    \\
        HRLLC $(D_k,C_k,\tau_s^{\max},\epsilon_s)$                        & $(0.25~\mathrm{MB},\,0.25~\mathrm{GCy},\,0.08~\mathrm{s},\,0.10)$ \\
        eMBB $(D_k,C_k,\tau_s^{\max},\epsilon_s)$                         & $(2.2~\mathrm{MB},\,0.75~\mathrm{GCy},\,0.25~\mathrm{s},\,0.20)$  \\
        mMTC $(D_k,C_k,\tau_s^{\max},\epsilon_s)$                         & $(0.10~\mathrm{MB},\,0.15~\mathrm{GCy},\,0.45~\mathrm{s},\,0.25)$ \\
        Predict (LSTM/kernel,strde,pool)      & 128 units / (3, 2, 2) \\
        Learning rate / $\mu_{\mathrm{RL}}$ / $\lambda_{\mathrm{GAE}}$ & $10^{-4}$ / $0.99$ / $0.95$                                       \\
        PPO clip / entropy coeff.                                         & $0.2$ / $0.02$                                                    \\
        Train \& eval episode/Hidd. dim                                    & $1500$, $10$ / $128$                                                 \\
        GA Populate/Generate/Mutation                                     & $24$ / $25$ / $0.08$                                              \\
        \bottomrule
    \end{tabular}
    \vspace{-10pt}
\end{table}

\begin{figure*}[t!]\centering
    \vspace{-6pt}
    \includegraphics[width=7.01in]{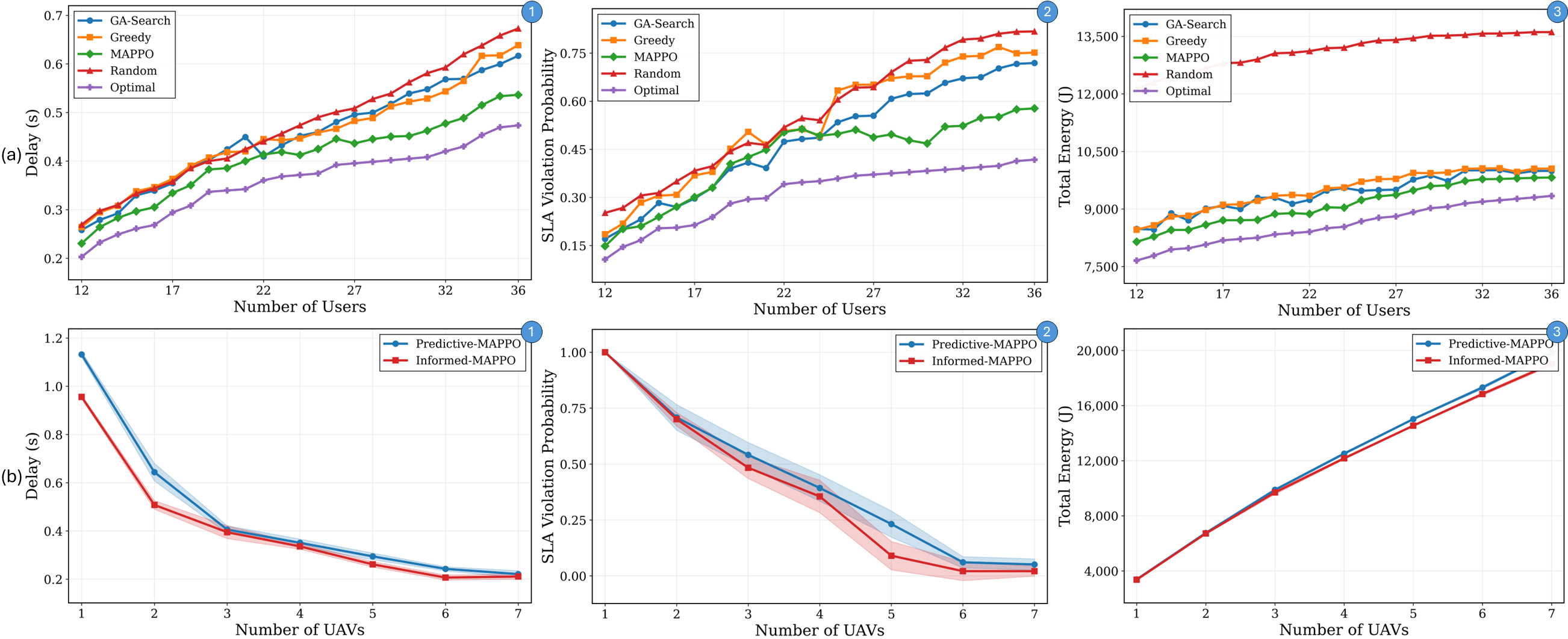}
    \vspace{-8pt}
    \caption{Performance under (a) increasing number of users and (b) increasing number of UAVs: (1) Delay, (2) SLA violation probability, and (3) UAV energy.}
    \label{figure:simulation_results}
    \vspace{-4pt}
\end{figure*}

In the first scenario, we evaluate the effectiveness of the \textit{MAPPO} module by providing the same predicted mobility and task-generation information to all non-oracle methods: (i) \textit{GA-Search}, a genetic search over discretized trajectory candidates; (ii) \textit{Greedy}, which prioritizes users according to a computation-delay urgency score; and (iii) \textit{Random}, which selects movement/allocation randomly. Therefore, the performance differences mainly reflect how each method exploits the predicted system state to guide UAV movement and compute allocation. We also include an \textit{Offline-Optimal} oracle-style benchmark with full future information to assess the optimality gap. As shown in Fig.~\ref{figure:simulation_results}(a), increasing the number of users, which emulates user spikes, increases transmission/computation contention, leading to higher delays and pushing slices into violating regimes more frequently (higher $\bar{P}_s^{\mathrm{viol}}$) and for longer periods (higher $\tilde{T}_s^{\mathrm{viol}}$). Although all methods degrade with user density, MAPPO remains the most \emph{SLA-stable} non-oracle method and stays closest to the oracle. This is because MAPPO learns a coordinated multi-UAV policy that jointly considers future user distribution, UAV energy states, and computation load, enabling proactive repositioning and SLA-aware CPU allocation. 
In contrast, Greedy is myopic, GA-Search is constrained by its discretized search space and finite search budget, and Random lacks SLA-aware control. 
Overall, MAPPO achieves lower delay and violation probability while remaining \emph{competitive} in energy consumption, as its learned policy avoids oscillatory movements and inefficient over-provisioning; the reduced frequency of violating regimes is consistent with shorter violation persistence.


In the second scenario, we evaluate the prediction module by comparing \textit{Predictive-MAPPO} with \textit{Informed-MAPPO}, where the latter uses full future mobility information and serves as an upper-bound reference for prediction quality. Fig.~\ref{figure:simulation_results}(b) shows that increasing the number of UAVs, which emulates resource sufficiency, significantly reduces delay and SLA violation probability for both methods by improving spatial coverage, shortening user-UAV distances, and increasing computation capacity. The gap between Predictive-MAPPO and Informed-MAPPO is more visible with fewer UAVs, as prediction errors are more harmful under scarce aerial resources and cause suboptimal repositioning or computation bottlenecks. However, as the number of UAVs increases, Predictive-MAPPO closely approaches Informed-MAPPO in delay and SLA violation probability, showing that the proposed predictor provides \textit{sufficiently accurate} future information for proactive slicing. Energy consumption increases for both methods as more UAVs participate in movement and computation, while their energy curves remain \textit{almost aligned}. This indicates that the gain of Informed-MAPPO mainly comes from more accurate anticipation rather than excessive energy use, confirming that Predictive-MAPPO achieves near-informed SLA-aware control using only learned predictions.

\section{Conclusion}\label{sec:conclusion}

In this paper, we studied SLA-aware network slicing for UAV-enabled MEC under user mobility, stochastic task arrivals, and limited onboard energy as well as computing resources. To address the resulting stochastic, non-convex, and time-coupled control problem, we proposed a predictive multi-agent framework in which cooperative UAV agents are trained with MAPPO under centralized training and decentralized execution, utilizing mobility and task-generation predictions to act proactively. Simulation results showed that the proposed Predictive-MAPPO improves SLA stability (lower violation probability and shorter violation duration) while remaining competitive in energy consumption and delay performance compared with baselines, and approaches the oracle benchmark with sufficiently accurate predicted information.
Future work will incorporate more realistic propulsion and interference models; treat uplink transmit power explicitly as an optimization variable for joint power, trajectory, association, and computation control; and consider dynamic slice admission control as well as adaptive bandwidth allocation. Moreover, we will explore LLM-driven \emph{agentic} orchestration~\cite{11103499} and semantic-aware control~\cite{11140421} for UAV-enabled slicing, combining high-level planning with continual learning and semantics-oriented reward feedback.

\section*{Acknowledgment}
The research work is supported in part by the Federal Ministry of Research, Technology, and Space (BMFTR), Germany, through the Project 6GEM+ under Grant 16KIS2411; the European Union’s Horizon Europe research and innovation programme under the 6G-Path project (Grant No. 101139172); and the Research Council of Finland 6G Flagship Programme under Grant No. 369116.

\bibliographystyle{IEEEtran}
\bibliography{References/conf_short, References/IEEEabrv, References/ref}

@STRING{IEEE_J_VT         = "{IEEE} Trans. Veh. Technol."}

@STRING{IEEE_J_JSAC       = "{IEEE} J. Sel. Areas Commun."}

@STRING{IEEE_J_WCOM       = "{IEEE} Trans. Wireless Commun."}

@STRING{IEEE_J_NSM        = "{IEEE} Trans. Netw. Service Manag."}

@STRING{IEEE_J_CCN        = "{IEEE} Trans. on Cogn. Commun. Netw."}

@STRING{IEEE_M_COM        = "{IEEE} Commun. Mag."}

@STRING{IEEE_M_NET        = "{IEEE} Netw."}

@STRING{IEEE_M_WC         = "{IEEE} Wireless Commun."}

@string{ wcnc = {Proc. IEEE Wireless Commun. and Networking Conf.}}

@inproceedings{farhoudi2025deep,
  title        = {Deep learning based service composition in integrated aerial-terrestrial networks},
  author       = {Farhoudi, Mohammad and others},
  booktitle    = {Proc. {IEEE} Int. Conf. Netw. Softwarization},
  pages        = {204--208},
  year         = {2025},
  organization = {IEEE}
}

@article{faraci2020design,
  title     = {Design of a {5G} network slice extension with {MEC UAVs} managed with reinforcement learning},
  author    = {Faraci, Giuseppe and Grasso, Christian and Schembra, Giovanni},
  journal   = IEEE_J_JSAC,
  volume    = {38},
  number    = {10},
  pages     = {2356--2371},
  year      = {2020},
  publisher = {IEEE}
}

@ARTICLE{farhoudi2026energy,
  author={Farhoudi, Mohammad and Mazandarani, Hamidreza and Shokrnezhad, Masoud and Taleb, Tarik and Lacalle, Ignacio},
  journal=IEEE_J_VT, 
  title     = {Energy Efficient Orchestration in Multiple-Access Vehicular Aerial-Terrestrial {6G} Networks},
  year={2026},
  volume={},
  number={},
  pages={1-16},
  doi={10.1109/TVT.2026.3682050}}

@article{wu2023intelligent,
  title     = {Intelligent and survivable resource slicing for {6G}-oriented {UAV}-assisted edge computing networks},
  author    = {Wu, Guoquan and Zhang, Bing and Li, Ya},
  journal   = {Computer Commun.},
  volume    = {202},
  pages     = {154--165},
  year      = {2023},
  publisher = {Elsevier}
}

@article{chen2025qos,
  title     = {{QoS}-oriented task offloading in {NOMA}-based Multi-{UAV} cooperative {MEC} systems},
  author    = {Chen, Peipei and others},
  journal   = IEEE_J_WCOM,
  year      = {2025},
  publisher = {IEEE}
}

@article{li2025self,
  title     = {Self-adjusting network slicing for dynamic heterogeneous task offloading in {UAV}-enabled mobile edge computing},
  author    = {Li, Xulong and others},
  journal   = IEEE_J_CCN,
  year={2026},
  volume={12},
  number={},
  pages={673-687},
  publisher = {IEEE}
}

@article{tian2023service,
  title     = {Service satisfaction-oriented task offloading and {UAV} scheduling in {UAV}-enabled {MEC} networks},
  author    = {Tian, Jie and Wang, Di and Zhang, Haixia and Wu, Dalei},
  journal   = IEEE_J_WCOM,
  volume    = {22},
  number    = {12},
  pages     = {8949--8964},
  year      = {2023},
  publisher = {IEEE}
}

@article{tang2022slicing,
  title     = {Slicing-based software-defined mobile edge computing in the air},
  author    = {Tang, Jianhang and Nie, Jiangtian and Zhao, Jun and Zhou, Yi and Xiong, Zehui and Guizani, Mohsen},
  journal   = IEEE_M_WC,
  volume    = {29},
  number    = {1},
  pages     = {119--125},
  year      = {2022},
  publisher = {IEEE}
}

@techreport{ITUR-M2160,
  author      = {{ITU-R}},
  title       = {Framework and overall objectives of the future development of {IMT} for 2030 and beyond},
  institution = {International Telecommunication Union, Radiocommunication Sector},
  number      = {{M.2160-0}},
  month       = nov,
  year        = {2023}
}

@inproceedings{sasan2024joint,
  title        = {Joint network slicing, routing, and in-network computing for energy-efficient {6G}},
  author       = {Sasan, Zeinab and others},
  booktitle    = wcnc,
  pages        = {1--6},
  year         = {2024},
  organization = {IEEE}
}

@article{sasan2025balancing,
  title     = {Balancing resource utilization and slice dissatisfaction through dynamic soft slicing for {6G} wireless networks},
  author    = {Sasan, Zeinab and Khorsandi, Siavash},
  journal   = {Scientific Reports},
  volume    = {15},
  number    = {1},
  pages     = {22987},
  year      = {2025},
  publisher = {Nature Publishing Group UK London}
}

@article{yabe2024yjmob100k,
  title     = {{YJMob100K}: City-scale and longitudinal dataset of anonymized human mobility trajectories},
  author    = {Yabe, Takahiro and others},
  journal   = {Scientific Data},
  volume    = {11},
  number    = {1},
  pages     = {397},
  year      = {2024},
  publisher = {Nature Publishing Group UK London}
}

@inproceedings{11140421,
  author    = {Mazandarani, Hamidreza and others},
  booktitle = {Proc. Int. Conf. Mach. Learn. for Commun. and Netw.},
  title     = {Semantic-Aware Dynamic and Distributed Power Allocation: a Multi-{UAV} Area Coverage Use Case},
  year      = {2025},
  volume    = {},
  number    = {},
  pages     = {1-6},
  doi       = {10.1109/ICMLCN64995.2025.11140421},
}

@article{11103499,
  author   = {Shokrnezhad, Masoud and Taleb, Tarik},
  journal  = IEEE_M_COM,
  title    = {An Autonomous Network Orchestration Framework Integrating Large Language Models with Continual Reinforcement Learning},
  year     = {2025},
  volume   = {63},
  number   = {8},
  pages    = {78-84},
  doi      = {10.1109/MCOM.001.2400526},
}

@article{farhoudi_discovery_2025,
  author       = {Mohammad Farhoudi and Masoud Shokrnezhad and Tarik Taleb and Richard Li and JaeSeung Song},
  title        = {Discovery of {6G} Services and Resources in Edge-Cloud-Continuum},
  journal      = IEEE_M_NET,
  volume       = {39},
  number       = {3},
  pages        = {223--232},
  year         = {2025},
  doi          = {10.1109/MNET.2024.3438096}
}

@misc{yu2022MAPPO,
      title={The Surprising Effectiveness of {PPO} in Cooperative, Multi-Agent Games}, 
      author={Chao Yu and others},
      year={2022},
      eprint={2103.01955},
      archivePrefix={arXiv},
      primaryClass={cs.LG},
      url={https://arxiv.org/abs/2103.01955}, 
}

@ARTICLE{10855598,
  author={Barick, Subhrajit and Singhal, Chetna},
  journal=IEEE_J_NSM, 
  title={{UAV}-Assisted {MEC} Architecture for Collaborative Task Offloading in Urban {IoT} Environment}, 
  year={2025},
  volume={22},
  number={1},
  pages={732-743},
  doi={10.1109/TNSM.2025.3535094}}

\end{document}